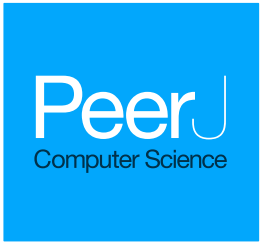

# Increasing the degree of parallelism using speculative execution in task-based runtime systems

Bérenger Bramas

CAMUS Team, Inria Nancy—Grand Est, Illkirch-Graffenstaden, France

## ABSTRACT

Task-based programming models have demonstrated their efficiency in the development of scientific applications on modern high-performance platforms. They allow delegation of the management of parallelization to the runtime system (RS), which is in charge of the data coherency, the scheduling, and the assignment of the work to the computational units. However, some applications have a limited degree of parallelism such that no matter how efficient the RS implementation, they may not scale on modern multicore CPUs. In this paper, we propose using speculation to unleash the parallelism when it is uncertain if some tasks will modify data, and we formalize a new methodology to enable speculative execution in a graph of tasks. This description is partially implemented in our new C++ RS called SPETABARU, which is capable of executing tasks in advance if some others are not certain to modify the data. We study the behavior of our approach to compute Monte Carlo and replica exchange Monte Carlo simulations.

## INTRODUCTION

Parallel CPUs are now everywhere, from mobile phones to high-performance computing nodes. To efficiently use this type of architecture, it is necessary to design applications to execute in parallel. This parallelization can be done in many ways with different paradigms. Among them, the task-based approaches have demonstrated successful exploitation of the parallelism of an algorithm by simply using the *real* dependencies between the data. In turn, the division of the algorithm into tasks cannot use dependencies at a low level, such as the instruction level, because of the runtime and context management overheads but has to instead find the appropriate granularity to balance between the degree of parallelism and chunk of work. Using task-based methods, it is possible to dissociate the parallelization algorithm from the rest of the code, composed of computational kernels and data structure definitions, but also from the hardware. This is intended to reduce issues coming from management of parallelization but also to avoid re-writing the application due to hardware changes. The complexity is then in the runtime system (RS), which is in charge of managing the execution, the distribution of the work, and the coherency. This also gives opportunities for specialists in scheduling or concurrent programming to improve these layers with a possible direct impact on the applications on the top.

On the other hand, the features and expressiveness of each RS can be different, and some very specific features could be needed for certain classes of algorithms. This was the starting point of our work while developing Monte Carlo (MC) and replica exchange Monte Carlo (REMC) algorithms for protein simulations. The description using tasks and dependencies of these algorithms gives a low degree of parallelism. However, some *write* dependencies are actually not true because a task might ultimately not modify the data, but this cannot be known in advance. This motivated us to look to the side of speculative execution, with a primary objective of improving the scalability of the MC/REMC algorithms. Secondary objectives are to get a generic pattern/method to use speculation in task-based RS's and to use only features that already exist in most task-based RS's (tasks).

The main contributions of our study are the following:

- Describe a general approach to include speculation in task-based RS's.
- Detail a possible implementation of this new strategy.
- Introduce SPETABARU, a new task-based RS capable of speculation.
- Illustrate how speculation can speed up MC and REMC simulations.

The current paper is organized as follows. In 'Motivation Example: Monte Carlo Simulations', we describe the MC and REMC simulations and discuss how they are usually parallelized as a motivation to our work. 'Background' introduces the notions related to RS's and speculative executions. 'Speculation in Task-Based Runtime Systems' describes our new strategy and how it can be implemented in a regular task-based RS. Finally, 'Performance Study' presents a performance study for executing MC and REMC simulations with speculation.

# MOTIVATION EXAMPLE: MONTE CARLO SIMULATIONS

The MC method is widely used in simulations of physical systems, especially when there are many coupled degrees of freedom that make traditional approaches inefficient or too expensive. Among this large class of solvers, we focus on the Monte Carlo simulations that use the Metropolis–Hastings update. We refer to the studies in protein simulation (see *Thachuk, Shmygelska & Hoos, 2007*; *Kim & Hummer, 2008*) to illustrate more precisely the type of problem we focus on.

Let us consider a system composed of multiple groups of beads/particles called domains. A domain can be a peptide in studies that focus on biomolecules in an all-atom approach. The energy of the system is a quadratic pair-wise computation between all particles in the system. A domain can move, that is, it can rotate, shift, or even redistribute its particles in space, which leads to a recomputation of the energy. The objective of such a problem generally is to find the configuration, meaning the position of the domains/beads, with the lowest energy.

In the corresponding algorithm, a single stochastic process evaluates the energy of the system and accepts/rejects updates based on the temperature $T$. This $T$ influences the probability for a change to be accepted, and with a high temperature the updates are more likely to be accepted. The MC simulation method is shown in Algorithm 1. At line 3, the energy is computed for the default configuration. Then, for a given number of iterations, the algorithm uses the domains and updates them (see line 10). It continues by computing the new energy with this domain that moved at line 11. At line 13, we use the Metropolis formula to decide, based on the energy difference and temperature, if the change has to be accepted. In case it is, we keep the change by replacing the old domain's position with the new one (see line 14).

---

**ALGORITHM 1:** Monte Carlo simulation algorithm.

---

1  **function** *MC(domains, temperature)*
2     `// Compute energy (particle to particle interactions)`
3     energy ← compute_energy(domains)
4     MC_Core(domains, temperature, energy)
5  **function** *MC_Core(domains, temperature, energy)*
6     `// Iterate for a given number of iterations`
7     **for** *iter* **from** *1* **to** *NB_LOOPS_MC* **do**
8         **for** *d in domains* **do**
9             `// Move a domain and compute new energy`
10             new_d ← move(temperature, d)
11             new_energy ← update_energy(energy, new_d, domains)
12             `// Accept the move (or do nothing)`
13             **if** *random_01() ≤ metropolis(new_energy, energy, temperature)* **then**
14                domains ← replace(domains, d, new_d)
15                energy ← new_energy
16             **end**
17         **end**
18     **end**

---

### Replica exchange Monte Carlo simulation (REMC)

The MC algorithm might get trapped in local minimums or miss all local minimums, depending on $T$. The idea of the REMC, also known as parallel tempering, is to run several MC simulations of the same system, but each with a different temperature. Consequently, the acceptance rate and the speed of changes are very different for each of them. Then, at defined iterations, the simulations are exchanged, again using the Metropolis formula. Therefore, simulations that run at high temperatures and where modifications were easily accepted will then run at a lower temperature.

Algorithm 2 contains the different steps to be done with $N$ different replicas/temperatures. At line 4, we compute the energy for all randomly initialized systems. Then, the algorithm iterates for a given number of times and first performs a call to the MC algorithm for each configuration, line 10. Second is an exchange stage between configurations, starting at line 7. It is common that the *exchange_list* is designed such that the exchange happens between odd–even pairs of simulation when *iter* is odd and even–odd otherwise.

### Parallelization of MC and REMC

The data dependencies for both the MC and REMC algorithms are easy to extract.

In the MC algorithm, each iteration depends on the previous one. The changes are only dependent over the domain, but to compute a new energy we have to know if the change of the previous domain has been accepted or not. Consequently, the only parallelism that can be applied is inside the computation of the energy, possibly with a fork-join strategy.

The same is true for the REMC, with the difference that the calls to the $N$ different MC simulations can be done in parallel. Therefore, it is common to use one thread or one process per replica. In *Altekar et al. (2004)*, the authors proposed a point-to-point exchange

---

**ALGORITHM 2:** Replica Exchange Monte Carlo (parallel tempering) simulation algorithm.

```
1  function REMC(domains[N], temperature[N])
2  |  // Compute energy (particle to particle interactions)
3  |  for s from 1 to N do
4  |  |  energy[s] ← compute_energy(domains[s])
5  |  end
6  |  // Iterate for a given number of iterations
7  |  for iter from 1 to NB_LOOPS_REMC do
8  |  |  for s from 1 to N do
9  |  |  |  // Compute usual MC for each simulation
10 |  |  |  MC_Core(domains[s], temperature[s], energy[s])
11 |  |  end
12 |  |  // Compare based on a given strategy
13 |  |  for s in exchange_list(iter) do
14 |  |  |  // Use the energy difference between s and s+1 to decide to
        |  |  |      exchange them
15 |  |  |  if random_01() ≤ metropolis(energy[s] - energy[s+1], temperatures[s]) then
16 |  |  |  |  swap(domains[s], domains[s+1])
17 |  |  |  |  swap(energy[s], energy[s+1])
18 |  |  |  end
19 |  |  end
20 |  end
```

---

scheme. They distributed N replicas among P processes and ensured that there was no global barrier; in other words, only the processes that have to exchange simulations communicate. The same principle applies in *Zhou et al. (2013)* with one process per temperature/replica. Similarly, in *Gross, Janke & Bachmann (2011)*, the authors dedicated one thread or one GPU per replica. Finally, in *Treikalis et al. (2016)*, the authors proposed a parallel framework for plugging in MC-based applications. They also remind that asynchronous RE (so without a global barrier) is needed in many cases.

However, no matter how efficient the implementation, synchronizations between the domains in the MC and the replicas in the REMC must be done. But it appears clear that these dependencies are sometimes not needed when the changes are rejected because the data are left unchanged, so it could be possible to compute in advance, hoping that the result will not be invalid. The main part of the current study is to provide a system for this compute-in-advance on top of an RS.

# BACKGROUND

## Task-based parallelization

Most HPC applications that support shared-memory parallelization rely on a fork-join strategy. Such a scheme uses a simple division of the work into independent operations and a global barrier that ensures all the work is done before the execution continues. The most common feature is the for-loop parallelization using pragmas as proposed by the OpenMP (*OpenMP Architecture Review Board, 1997*) standard. This model has been extended to a tasks-and-wait scheme, where operations come from different parts of the application but are still independent. The task model from OpenMP 3 (*OpenMP Architecture Review Board, 2008*; *Ayguadé et al., 2009*) and the task-based programming language Cilk (*Blumofe et al., 1996*) (later extended in Cilk++ (*Leiserson, 2009*) and Cilk Plus (*Intel, 2017*)) follow this idea. This is still a fork-join model because successive spawn phases of independent tasks (fork) must be explicitly synchronized (join) to ensure a correct execution. This really limits the scalability because of the waiting time and the imbalance between tasks. Developers are able to increase the degree of parallelism by using multiple sources of tasks that they known are independent. But then it starts to become manual management of the dependencies, which a modern task-based RS is intended to do.

An algorithm can be decomposed in interdependent operations where the output of some tasks is the input of others. A task-based implementation will map tasks over these operations and create dependencies between them to ensure execution coherency. The result can be seen as a direct acyclic graph (DAG) of tasks, or simple graph of tasks, where each node is a task and each edge is a dependency. An execution of such a graph will start from the nodes that have no predecessor and continue inside the graph, ensuring that when a task starts, all its predecessors have completed. The granularity of the tasks, that is, the content in terms of computation, cannot be too fine-grained because the internal management of the graph implies an overhead that must be negligible to ensure good performance, as shown in *Tagliavini, Cesarini & Marongiu (2018)*. Therefore, it is usually the developer's responsibility to decide what a task should represent. The granularity is then a balance between the degree of parallelism and the RS overhead. For that reason, several researches are conducted to delegate partially or totally the RS system to the hardware with the objective of relieving the worker threads, as in *Chronaki et al. (2018)*.

Building a graph of tasks can be done in two major ways. The first possibility is to build a graph by creating the nodes and connections between them explicitly, as it is used in the parametrized task graph (PTG) model (*Cosnard & Loi, 1995*). This approach is complex and usually requires completely rewriting an application. The second method is the sequential task flow (STF) (*Agullo et al., 2016b*). Here, a single thread creates the tasks by informing the RS about the access of each of them on the data. The RS is then able to generate the graph and guarantee that the parallel execution will have the absolute same result as a sequential one. This ends in a very compact code with few modifications required to add to an existing application by moving the complexity in the RS. In our work, we use the STF model.

There now exist numerous different task-based RS's. The most popular ones are implementations of the OpenMP version 4 (*OpenMP Architecture Review Board, 2013*) standard that defines the additional pragma keyword *depend* to inform the RS about the type of data accesses performed by the tasks. However, using pragmas, in general, is tedious when a task has hundreds of dependencies or when the number of dependencies are known at runtime, because it ends up being an ugly and error-prone code. In addition, as OpenMP is a standard, it is upgraded slowly to ensure backward compatibility. Moreover, the standard is weak in the sense that it does not impose any constraints on the implementation and complexity of the underlying algorithms. This can cause performance surprises for the user when compared to different OpenMP RS's. Nonetheless, its portability, stability, and maturity make it a safe long-term choice.

Additionally, some industrial RS's have been defined, such as the Intel Threading Building Blocks (ITBB), C++ library (*Intel, 2017b*). It allows building of graphs using the PTG model, but it also supports various other features such as memory allocation management and parallel containers. PaRSEC (*Danalis et al., 2014*) is another RS based on the PTF model that had been demonstrated to be effective in various scientific applications. CHARM++ (*Kale & Krishnan, 1993*) is a C++-based parallel programming system. It includes the parallelism by design with a migratable-objects programming model, and it supports task-based execution. SMPSs (*Perez, Badia & Labarta, 2008*), now included in OmpSs (*Duran et al., 2011*), is a RS based on the STF model that defined what later became the OpenMP tasks. It uses pragmas annotation to inform the RS about data access by the tasks. StarPU (*Augonnet et al., 2011*) is an RS that was first designed to manage heterogeneous architectures. It is a C library such that the user has to use low-level programming and function pointers. However, it is extremely flexible and it is capable of having distributed memory STF. XKaapi (*Gautier et al., 2013*) is an RS that can be used with standard C++ or with specific annotation but which needs a given compiler. Legion (*Bauer et al., 2012*) is a data-centric programming language that allows for parallelization with a task-based approach. SuperGlue (*Tillenius, 2015*) is lightweight C++ task-based RS. It manages the dependencies between tasks using a data version pattern.

Most of these tools support a core part of task-based RS, such as creating a graph of tasks (even if it is implemented differently) where tasks can read or write data. However, scheduling is an important factor in the performance (*Agullo et al., 2016a*), and few of these RS's propose a way to create a scheduler easily without having to go inside the RS's code. Moreover, specific features provide mechanisms to increase the degree of parallelism. For instance, some RS's allow specification of whether data access is commutative, meaning that the tasks write data but the order is not important. This kind of advanced functions can make a big difference in terms of performance (*Agullo et al., 2017*). We refer to *Thoman et al. (2018)* for a more in-depth comparison of the RS's.

To the best of our knowledge, none of them propose a speculation system. But any of the STF based RS could easily implement our strategy to use speculation.

## Speculative execution

Speculation is the principle of guessing without being certain in order to make a profit. It is commonly used at the instruction level in CPU hardware to fill the pipeline of instruction when there is a branch, to prefetch the memory, to manage the memory dependence, and to use transactional memory. In our system, the speculation is at the application level, or more precisely, at the task level. Therefore, it has very different constraints, advantages, and disadvantages compared to hardware-based speculation. On the other hand, it requires managing copies and synchronizations at a high level, too.

The two main speculation strategies are called eager and predictive. Eager execution computes all paths, and this is usually not realistic because there are too many of them and because the real path may actually be found only once the results are known. Predictive execution is the more common strategy; in this model, the execution follows a path until we know if the prediction is true.

A common speculation pattern is called thread level speculation (TLS). In *Steffan & Mowry (1998)*, the authors describe how TLS was expected to improve parallel performance of non-numeric applications. The idea is to mimic the instruction speculation by the compiler by, for example, reordering instruction to move a *load* ahead of a *store*. At execution time, if the speculation appears unsafe, a recovery method should be used. TLS is somewhat similar, but the *load* and *store* are performed by two different threads. A speculation is then safe if the execution of a *load* does not target a location that will be modified by a *store*, which should have already been done. The main problems are to provide a system to control the safety at low costs and how the compiler can automatically insert speculative code. Consequently, many researchers have defined new hardware that would make the TLS or a similar pattern possible, such as in *Jeffrey et al. (2015)*. But many of these research projects are tied to the manufacturer's choices.

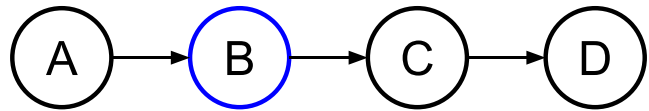

**Figure 1** Graph of four tasks where *B* is task that will potentially modify the data.

Full-size 🖼 DOI: 10.7717/peerjcs.183/fig-1

In *Salamanca, Amaral & Araujo (2017)*, the authors discuss how instructions that currently exist in most CPUs to manage transactional memory (TM) could be used to implement TLS. They show that TLS is already possible but that false-sharing, coming from the size of the element the TM instructions are working on, can significantly decrease the performance.

Other speculation methods have been proposed with pure software approaches. For example, the APOLLO (*APOLLO, 2016*) framework is capable of speculative parallelization processes for loops of any kind (for, while, or do-while loops). It is composed of two main layers. First, an extensions to the CLANG-LLVM compiler prepares the program by generating several versions of each target loop nest and several code snippets called *code bones* (*Martinez Caamaño et al., 2017*). During the execution, the RS orchestrates the execution of the different code versions by chunks.

Our approach is different; because it is high level and not designed for a few instructions, it does not require special hardware instruction, and it is designed for applications that are already parallel at the top of a graph of tasks.

## SPECULATION IN TASK-BASED RUNTIME SYSTEMS
### Description
Consider a simple graph of tasks as shown in Fig. 1 where four tasks access the same data. Here, task *B* may or may not write data, but to ensure a correct execution, we must use *WRITE* data access, and so the dependency between *B* and *C* is strict.

In our approach, we allow the programmer to indicate that a task will potentially *WRITE* data, that is, there is a condition in the task that will make the modification effective or not. Then, at the end of the task's execution, this one informs the RS to detail whether or not the data was modified. In doing this, the RS knows that speculation is possible, and it can create extra tasks as shown in Fig. 2. When an *uncertain* task (a task with at least one potential *WRITE*) is inserted, the RS creates a copy-task in front of it, a *speculative task* and a *select task*. At runtime, if the speculation is enabled, the RS has to manage them to

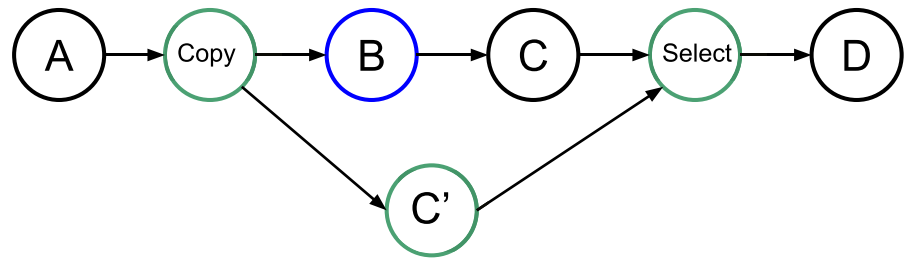

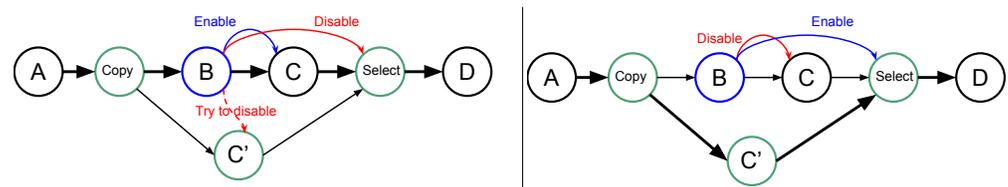

a: If B wrote on the data, the RS tries to cancel C', it
enables C and disables the select.

b: If B did not write on the data, the RS disables C and
enables the select.

ensure accuracy and coherency. In our example, when the speculation is enabled, both B
and C' can be computed concurrently. Then, there are two possibilities, either B modified
the data or it did not. If it did, as shown by Fig. 3A, then the result of C' is invalid and will
be destroyed. In addition, the valid result from B used by C will have to be computed, and
we can disable the select task. Otherwise, as shown in Fig. 3B, in the case where B did not
write data, then the output of C' is valid. C is disabled, and the select task is enabled to
ensure that the valid result is used as output of this group of tasks.

In terms of execution time, without speculation, we have the total duration D = D(B)
+ D(C). With speculation and if B writes data, then D = D(copy) + Max(D(B) + D(C),
D(C')), considering that C' is computed at the same time as B and C and where D(C')
is zero if canceled. With speculation and if B does not write data, then D = D(copy) +
Max(D(B),D(C')) + D(select).

The creation of the extra tasks can be done at the time of insertion, and thus, it does
not require any specific feature from the RS. However, it implies a possible overhead since
the creation of a task usually requires multiple allocations and possible synchronizations
in the scheduler. The enabling and disabling of the tasks during execution do not mean
that the task has to be removed from the DAG but that their core part should act as an
empty function. The select tasks are functions to act as a switch and to decide which data
are the valid output. For instance, in the given example, if $x$ is the original value accessed
by the tasks and $x'$ the copy of $x$ accessed by C', the select code would be $x = x'$. Therefore,
we only have to enable the select task when we want to overwrite $x$ by the output of C'.

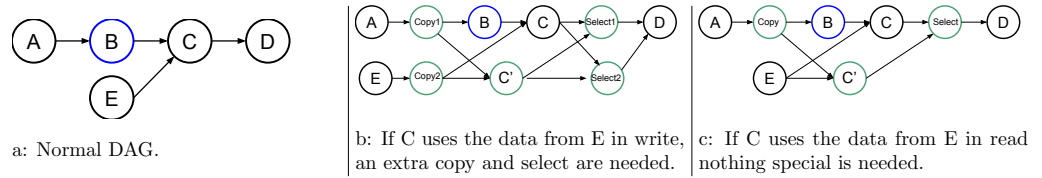

a: Normal DAG.

b: If C uses the data from E in write, an extra copy and select are needed.

c: If C uses the data from E in read nothing special is needed.

**Figure 4** Graph of five tasks where *B* is an uncertain task and where C, the task use for speculation, uses data from another task E.

Full-size 🖾 DOI: 10.7717/peerjcs.183/fig-4

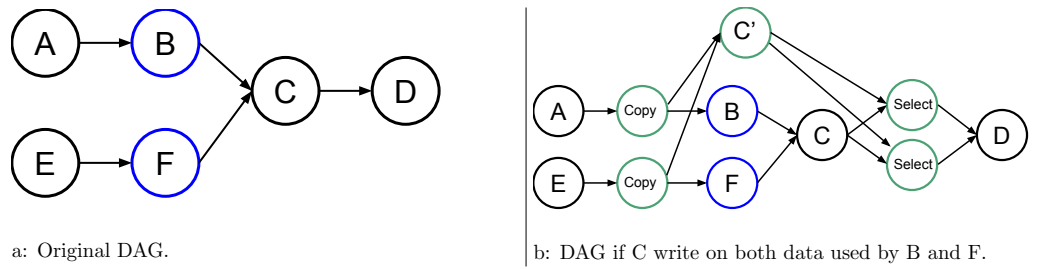

a: Original DAG.

b: DAG if C write on both data used by B and F.

**Figure 5** Graph of six tasks where *B* and *F* are non consecutive uncertain tasks.

Full-size 🖾 DOI: 10.7717/peerjcs.183/fig-5

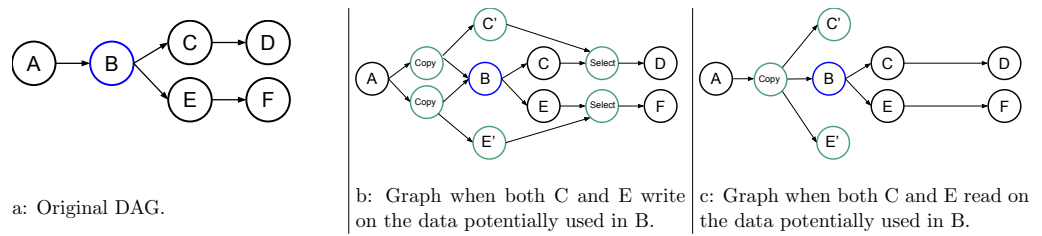

a: Original DAG.

b: Graph when both C and E write on the data potentially used in B.

c: Graph when both C and E read on the data potentially used in B.

**Figure 6** Graph of six tasks where *B* is an uncertain task.

Full-size 🖾 DOI: 10.7717/peerjcs.183/fig-6

The decision to activate the speculation can be done at runtime, and the extra tasks should simply be disabled if the speculation should not happen.

## Multiple dependencies

More advanced examples are shown in Figs. 4, 5, and 6. They show how the RS should act to speculate but still have valid and coherent execution.

In Fig. 4, task C uses data from a normal task E in addition to that from the uncertain task B. Therefore, if C writes data from E, C' does so as well. Since we do not know if C' will be valid, we have to insert an extra copy, as shown in Fig. 4B. An additional select is also needed to select the correct data from C or C' as the output of the task group. Otherwise, if the data from E is used in reading by C, then both C and C' can use the same data concurrently, as shown in Fig. 4C.

## Speculative task group (STG)

In Fig. 5, we provide an example where we have more than one uncertain task but not consecutively. In this case, we need to copy all data that could be modified by the uncertain tasks B and F, and we need one select for each of them. However, there is a very important aspect here because B and F might not have the same behavior, and as a result, one could write the data while the other does not. Therefore, in our approach, we link together all the tasks that are connected inside an STG. If at least one uncertain task of the group modifies the data, then the speculation has failed no matter the result of the other uncertain tasks. It could be more fine-grained, but that would require a complex algorithm to ensure coherent progression in the DAG. In the given example, if B or F modifies the data, then the RS tries to cancel C', enable C, and disable the selects.

An STG also links the tasks used for speculation. As an example, in Fig. 6, the uncertain task B may or may not write data at two different points. One of the two is later used by C, and the other one is later used by E. Here, we consider only two cases; when C and E write data, see Fig. 6B, and when they use those data in read, Fig. 6C is used. In the first case, when B fails, the RS tries to cancel C' and E', enable C and E, and disable the select, while in the second case, there is no need to have a select.

In our case, an STG is composed of several lists: the list of copies tasks, the list of uncertain tasks, the list of original speculative tasks, the list of speculative tasks, and the list of selects. At runtime, based on the result, the lists have to be properly managed to ensure a correct execution. The decision to enable the speculation in a task group can be done when the first copy task is ready. Since the construction of all this information is done at task insertion time, there is also the need to merge different task groups, which is simply a merge of the list, or an extra list that contains the other groups.

## Multiple consecutive uncertain tasks

It could happen that several uncertain tasks are consecutive (directly connected in the graph). There are several ways of managing such a configuration, and Fig. 7 shows four of them. In Fig. 7A, we leave everything unchanged compared to the previous examples, but then it appears that D could speculate over C. In Fig. 7B, we show that we could perform the speculation above all the uncertain tasks. Therefore, here, we let the consecutive uncertain tasks compute one after the other. However, with this approach, the degree of parallelism remains two no matter the number of uncertain tasks.

In Fig. 7C, we mix the speculation above B and C. Here, the degree of parallelism is three because C', C, and D' can be computed concurrently. However, again, if there are more than three uncertain tasks, the degree will remain the same.

Finally, what we use is shown in Fig. 7D. Here, (B,C), C', and D' can be computed concurrently, and the degree of parallelism will increase with the number of uncertain tasks.

## Speedup for one or multiple consecutive uncertain tasks

The expected speedup is, of course, a matter of probability depending on the success rate of the speculation. For instance, let us consider that we have $N$ consecutive uncertain tasks

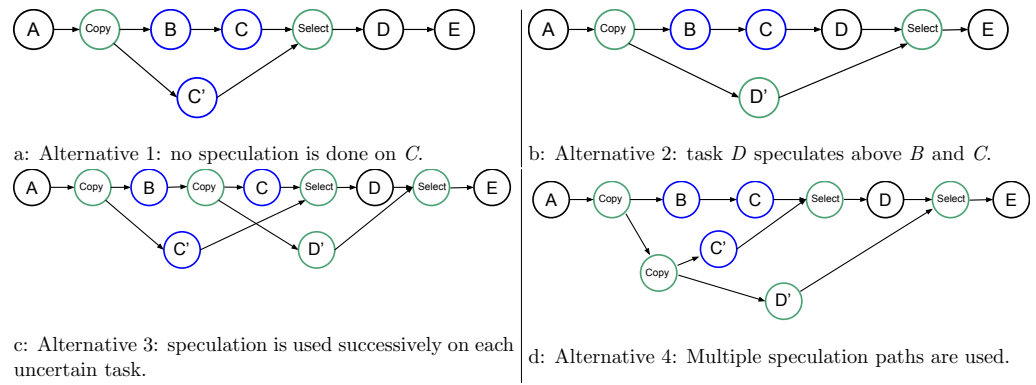

a: Alternative 1: no speculation is done on $C$.

b: Alternative 2: task $D$ speculates above $B$ and $C$.

c: Alternative 3: speculation is used successively on each uncertain task.

d: Alternative 4: Multiple speculation paths are used.

**Figure 7** Graph of four tasks where $B$ and $C$ are tasks that will potentially modify the data.
Full-size 🔍 DOI: 10.7717/peerjcs.183/fig-7

followed by a normal task all of the same cost $t$, negligible copies/selects, and at least $N$ available workers. The speedup will be, on average,

$$S = \frac{(N+1) \times t}{(N+1) \times t - D_N} \tag{1}$$

$$D_N = \sum_{i=1}^{N} \left( t \times i \times P_{i+1} \times \prod_{j=1}^{i} (1 - P_j) \right) \tag{2}$$

$$P_{N+1} = 1 \tag{3}$$

where $D_N$ is the average duration gain, and $P_i$ is the probability for the task of index $i$ to write data. In Eq. (2), we compute $D_N$ with a sum over the average gain when each of the uncertain tasks write data. For instance, when task $i+1$ writes data but all its predecessor tasks do not, we obtain an average gain of $i \times t$ with a probability of $P_{i+1} \times \prod_{j=1}^{j \leq i} (1 - P_j)$. In Eq. (3), $P_{N+1}$ is set to 1 because the definition specifies that the $(N+1)^{th}$ task is a normal task that always writes on the data.

If the probability is 1/2 for all uncertain tasks, then we have:

$$D_{1/2} = t \times \left( \sum_{i=1}^{N-1} \left( \frac{i}{2^{i+1}} \right) + \frac{N}{2^N} \right). \tag{4}$$

We provide the speedup in Table 1, if we consider an execution as a unit with $t = 1$ and we have $N$ uncertain tasks followed by one normal task.

## Multiple consecutive uncertain tasks (eager extension)

Our objective that is not yet implemented in our RS is presented in Fig. 8. Here, we create all the tasks necessary to be able to restart the speculation process when it fails. However, it requires the creation of $(N^2 + N)/2 - N$ speculative tasks, with $N$ being the number of consecutive uncertain tasks, plus the copy and select tasks. Many of these tasks will be disabled by default and enabled only when a speculation fails.

However, in terms of performance, any non-modification of the result provides a benefit of $t$. Therefore, the average duration gain $F(N)$ is given by

$$F(N) = F(N-1) \times P_N + (F(N-1)+t) \times (1-P_N) \qquad (5)$$

$$= F(N-1) + t \times (1-P_N) \qquad (6)$$

$$F(0) = 0. \qquad (7)$$

where $P_i$ is the probability for the task of index $i$ to write data. In Eq. (5), we compute $F(N)$ by summing first the gain when task $N$ writes data, which happens with a probability $P_N$, and second the gain when task $N$ does not write data, which happens with a probability $1-P_N$. The formula is recursive and can be rewritten by

$$F(N) = t \times \sum_{i=1}^{N} 1 - P_i. \qquad (8)$$

| $N$ | 1 | 2 | 3 | 4 | 5 | 6 | 7 |
|---|---|---|---|---|---|---|---|
| $D_{1/4}$ | 0.75 | 1.31 | 1.73 | 2.05 | 2.29 | 2.47 | 2.6 |
| $S_{1/4}$ | 1.6 | 1.78 | 1.77 | 1.7 | 1.62 | 1.54 | 1.48 |
| $D_{1/2}$ | 0.5 | 0.75 | 0.875 | 0.938 | 0.969 | 0.984 | 0.992 |
| $S_{1/2}$ | 1.33 | 1.33 | 1.28 | 1.23 | 1.19 | 1.16 | 1.14 |
| $D_{3/4}$ | 0.25 | 0.312 | 0.328 | 0.332 | 0.333 | 0.333 | 0.333 |
| $S_{3/4}$ | 1.14 | 1.12 | 1.09 | 1.07 | 1.06 | 1.05 | 1.04 |

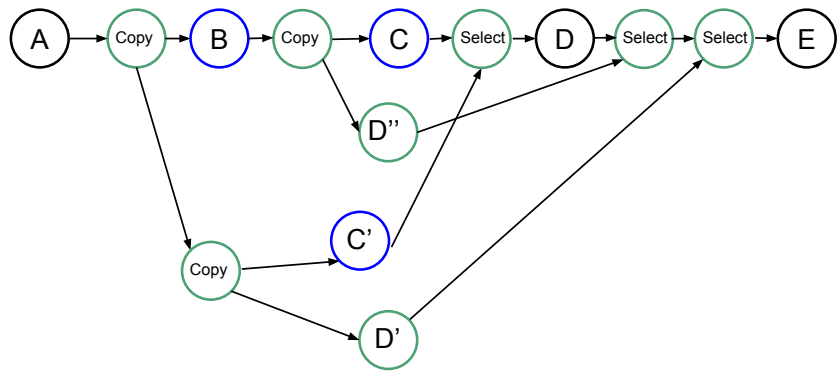

**Figure 8** Graph of four tasks where $B$ and $C$ are uncertain tasks. Extra tasks are created by the RS: C', D', D'' and several copies/selects.

Another way to describe this formula is to consider that if we have $N$ uncertain tasks, we first compute task 0 and speculate over it with duplicate of tasks from 1 to $N$. If tasks $i$ writes data, then we will have to compute task $i+1$ and speculate over it with duplicate of tasks from $i+2$ to $N$. We do so until no tasks write data or the $N^{th}$ is computed.

The average speedup is given by

$$S = \frac{(N+1) \times t}{(N+1) \times t - F(N)}. \qquad (9)$$

For a probability of 1/2 for all uncertain tasks, the average speedup is then equal to 2 no matter the number of consecutive speculative tasks.

## Changing the DAG on the fly

From the execution shown in Fig. 2, the question can be asked if the behavior could be different when C' is over before B. It would be interesting to speculate for D, too, by creating a task D' and the corresponding copy and select. However, we have decided not to do so because it requires modifying the DAG on the fly. This is a difficult operation to implement and one that most RS's do not support.

---

**ALGORITHM 3:** Uncertain task insertion.

```
 1  function insert_uncertain_task(t)
 2  │   // Clean the duplicate data handles if one is used in READ mode
    │       and it is used by t not in READ mode
 3  │   global_duplicates.clean_if_in_read(t.data in READ)
 4  │   // Find the spec-groups related to t, groups
 5  │   groups ← global_duplicates.find_spec_groups(t.data)
 6  │   if one of them is disabled or groups is empty then
 7  │   │   // Remove the duplicates related to t (not the one just
    │   │       inserted)
 8  │   │   global_duplicates.remove_any_kind(t.data)
 9  │   │   // Duplicate the data used by t in MAYBE-WRITE, list l1
10  │   │   l1 ← create_duplicate(t.data in MAYBE-WRITE)
11  │   │   // Insert t without speculation
12  │   │   internal_insert(t)
13  │   │   // Add l1 to the global list
14  │   │   global_duplicates.append(l1)
15  │   │   // Create a new group
16  │   │   g ← create_group_with_no_parent()
17  │   │   g.set_main_task(t)
18  │   end
19  │   else
20  │   │   // Duplicate the data used by t in MAYBE-WRITE, list l1
21  │   │   l1 ← create_duplicate(t.data in MAYBE-WRITE)
22  │   │   // Create a new group with groups as parents
23  │   │   g ← create_group_with_parents(groups)
24  │   │   g.add_copy_tasks(l1.copy_tasks)
25  │   │   // Duplicate the data used by t in MAYBE-WRITE that are already
    │   │       duplicated (list l1p) (inform g)
26  │   │   l1p ← create_duplicate(t.data in MAYBE-WRITE and exist in global_duplicates)
27  │   │   g.add_copy_tasks(l1p.copy_tasks)
28  │   │   // Duplicate the data used by t in WRITE that are not already
    │   │       duplicated (list l2) (inform g)
29  │   │   l2 ← create_duplicate(t.data in WRITE and do not exist in global_duplicates)
30  │   │   g.add_copy_tasks(l2.copy_tasks)
31  │   │   // Insert t as a normal task (inform g)
32  │   │   g.set_main_insert(t)
33  │   │   internal_insert(t)
34  │   │   // Insert t as a speculative task using data duplicates, l1 and
    │   │       l2 and the global list (inform g)
35  │   │   internal_speculative_insert(t, l1, l2)
36  │   │   // Add select tasks and clean duplicates (inform g)
37  │   │   select_tasks ← create_select_tasks()
38  │   │   g.add_select_tasks(select_tasks)
39  │   │   // Add l1p in the global list
40  │   │   global_duplicates.append(l1p)
41  │   end
```

---

## Algorithms
### Speculative task group
The STG contains lists of tasks that are all connected to the same uncertain tasks' results. It has a state, which can be undefined, enable or disable, and a result that indicates if one

---

**ALGORITHM 4:** Normal tasks insertion.

```
1  function insert_normal_task(t)
2      // Clean the duplicate data handles if one is used in READ mode
           and it is used by t not in READ mode
3      global_duplicates.clean_if_in_read(t.data in READ)
4      if no data used by t have been duplicated then
5          // Simply insert t, there is no speculation to do
6          internal_insert(t)
7      end
8      else
9          // Find the spec-groups groups related to t
10         groups ← global_duplicates.find_spec_groups(t.data)
11         if one of them is disabled then
12             // We already know that the speculation has failed
13             // Remove the duplicates related to t
14             global_duplicates.remove_any_kind(t.data)
15             // Insert t without speculation
16             internal_insert(t)
17         end
18         else
19             // Build a new group with groups as parents
20             g ← create_group_with_parents(groups)
21             // Duplicate the data used by t in WRITE mode that are not
                   already duplicate (list l2) (inform g)
22             l2 ← create_duplicate(t.data in WRITE and do not exist in global_duplicates)
23             // Insert t as a normal task (inform g)
24             g.set_main_task(t)
25             internal_insert(t)
26             // Insert t as a speculative task using data duplicates,
                   using l2 (inform g)
27             internal_speculative_insert(t,l1,l2)
28             // Add select tasks (use l2) and clean duplicates (inform g)
29             select_tasks ← create_select_tasks()
30             g.add_select_tasks(select_tasks)
31         end
32     end
```

---

of the uncertain tasks did write data. An STG also has a list of predecessor STGs and a list of successor STGs. In fact, if an STG is enabled and running and the uncertain task did not modify the data such that the speculation succeeded, then if a new tasks is inserted and uses data from this task group and another, they should be connected.

### Task insertion

In Algorithm 3, we give an overview of the task insertion algorithm of an uncertain task, and in Algorithm 4, we give that of a normal task. The algorithm uses a list called *global_duplicates* to save the duplicate data from the copy tasks inserted before the uncertain tasks. For instance, when a task is inserted, we can look up whether one of its dependencies has been duplicated or not. However, the difficulty arises from the construction at task insertion time, that is without a posteriori and without knowing what tasks will come next. Therefore, in order to have executions as defined in 'Description', we have to determine if there is a duplicate and if the related STGs have been disabled or failed. Also, we need to create a duplicate of the data but without using them for the current task, which is why we store the duplicate information in some other lists before adding them to the *global_duplicates* after the task has been inserted.

### At execution

During the execution, the RS has to decide if the speculation is enabled or not. It is convenient to do this when the first copy task of an STG becomes ready to be executed. In

                    

fact, the decision process can then use information such as the current number of ready tasks in the scheduler. Then, the tasks have to be enabled or disabled accordingly. After any uncertain task has finished, the RS has to enable or disable tasks related to the STG. It has to iterate over the lists of the STG and, potentially, on its successor. The DAG remains unchanged, and so any RS that has a callback and activation/deactivation features can implement our mechanism.

## Implementation (SPETABARU)

We have implemented our speculation method on top of a lightweight C++ RS that was originally less than 3,000 lines.[1] We use modern C++17 and advanced meta-programming to look at data dependencies at compile time. The RS supports the data access modes of *read*, *write*, *atomic_write*, and *commute* (for commutative operations). We also use a dynamic array view, which allows for having an unlimited number of known dependencies at execution time. It is possible to use lambda/anonymous functions as tasks, as shown in Code 1.

```cpp
// Create the runtime
const int NumThreads = SpUtils::DefaultNumThreads();
SpRuntime runtime(NumThreads);
const int initVal = 1;
int writeVal = 0;

// Create a task with lambda function
runtime.task(SpRead(initVal), SpWrite(writeVal),
            [](const int& initValParam, int& writeValParam){
    writeValParam += initValParam;
});
// Create a task with lambda function (that returns a bool)
auto returnValue = runtime.task(SpRead(initVal), SpWrite(writeVal),
                                [](const int& initValParam,
                                   int& writeValParam) -> bool {
    writeValParam += initValParam;
    return true;
});

// Wait completion of a single task
returnValue.wait();
// Get the value of the task
const bool res = returnValue.getValue();
// Wait until two tasks (or less) remain
runtime.waitRemain(2);
// Wait for all tasks to be done
runtime.waitAllTasks();
// Save trace and .dot
runtime.generateTrace("/tmp/basis-trace.svg");
runtime.generateDot("/tmp/basis-dag.dot");
```

Code 1: SPETABARU example.

The example of an uncertain task is given in Code 2. Compared to a regular task, the keyword *SpMaybeWrite* replaces *SpWrite* and the function returns a boolean to inform the RS if a modification occurs.

```cpp
runtime.potentialTask(SpMaybeWrite(val), [](int& /*valParam*/) -> bool {
    return false; // val has not been modified
});
```

Code 2: Example of creating an uncertain task in SPETABARU.

```
1  Task A : write(val)
2  Task B uncertain : maybe-write(val) -> false
3  Task C uncertain : maybe-write(val) -> true
4  Task D : write(val)
```

(a) Description of the tasks.

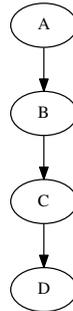

(b) DAG without speculation.

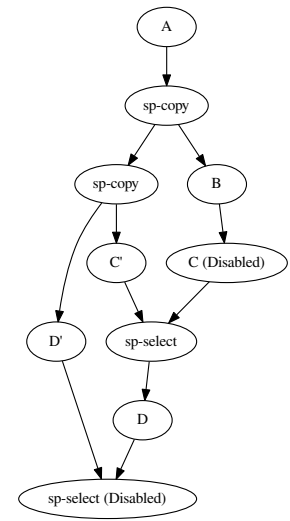

(c) DAG with speculation.

## Execution example

We show in Figs. 9 and 10 two examples and their execution results. The description of the tasks we insert are given in Figs. 9A and 10A, respectively. Those descriptions also indicate the result of the uncertain tasks known during execution at the end of the tasks, where true means that the task writes on the data and that otherwise, it did not. The DAGs without speculation are given in Figs. 9B and 10B, and the DAGs with speculation in Figs. 9C and 10C, respectively.

In Fig. 9C, after B is executed, the RS knows that B did not change any data. Therefore, it disables C and enables the merge. However, once C' is complete, the RS knows that it wrote data. Consequently, the RS must enable D and disable the merge. It also tried to disable D' but the result visible on the DAG indicates that this was too late.

In Fig. 10C, a more complex execution happened. Since B did not write data, C has been disabled and the corresponding select has been enabled. However, since D and E did write data, then F and G have been enabled, and the two last selects have been disabled.

## PERFORMANCE STUDY

### Configuration
#### *Software/Hardware*

We used an Intel Xeon E5-2680 v3 with a frequency of 2.5 GHz and 2 sockets of 12 cores each. We compiled using the GNU compiler version 7.2, and we bound the threads to the cores following a compact strategy, e.i. the threads are pinned to contiguous cores. We use SPETABARU available on the public master branch at tag *v0.15*.[2]

```
1  Task A : write(val1),write(val2),write(val3)
2  Task B uncertain : maybe-write(val1) -> false
3  Task C uncertain : maybe-write(val1) -> true
4  Task D uncertain : maybe-write(val2) -> true
5  Task E uncertain : maybe-write(val3) -> true
6  Task F : ...
       write(val1),write(val2),write(val3)
```

(a) Description of the tasks.

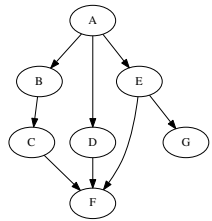

(b) DAG without speculation.

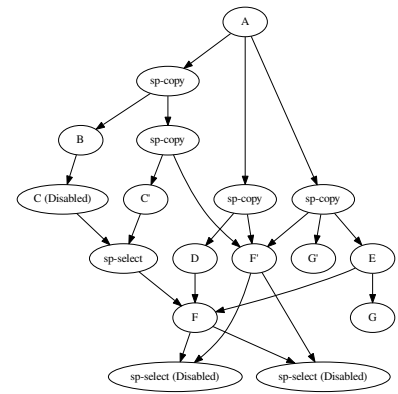

(c) DAG with speculation.

**Figure 10** Execution example with seven tasks. B, C, D and E are uncertain tasks and here B did not write on the data, while C, D and E did. Consequently, the RS has disabled or enabled the other tasks accordingly.

Full-size 🖼 DOI: 10.7717/peerjcs.183/fig-10

### Test case

We evaluated our approach on an MC simulation composed of five domains, where each domain had 2,000 particles. The reject/accept ratio was around 0.4, and we executed the simulation for 1 to 100 iterations. We computed the Lennard-Jones potential between the particles to obtain the global energy. The moves (particle update) are a simple random distribution of the particles in the simulation box. For the REMC, we used five replicas and performed a replica exchange every three iterations. The exchange rate between replicas was approximately 0.7.

Execution times are obtained by doing the average of 5 runs. The deviation is not given in the results because it is really limited (less than 1% in most cases). This is due to the kernels that are highly computational intensive with limited memory transfers, and because many threads remains idle for a significant amount of time during the execution, which help to get a limited contention effect.

## MC

We compared three approaches:

- Task-based: We used the task-based execution as baseline. However, since there is no degree of parallelism between the iteration of the loops it is equivalent as a sequential execution.
- Speculative *Spec(T,S)*: Here, *T* represents the number of threads and *S* the number of consecutive uncertain tasks inserted before inserting a normal task. The speculation is always enabled.
- Reject *Rej(T)*: Here, *T* represents the number of threads and *S* the number of consecutive uncertain tasks inserted before inserting a normal task. In this configuration, we looked

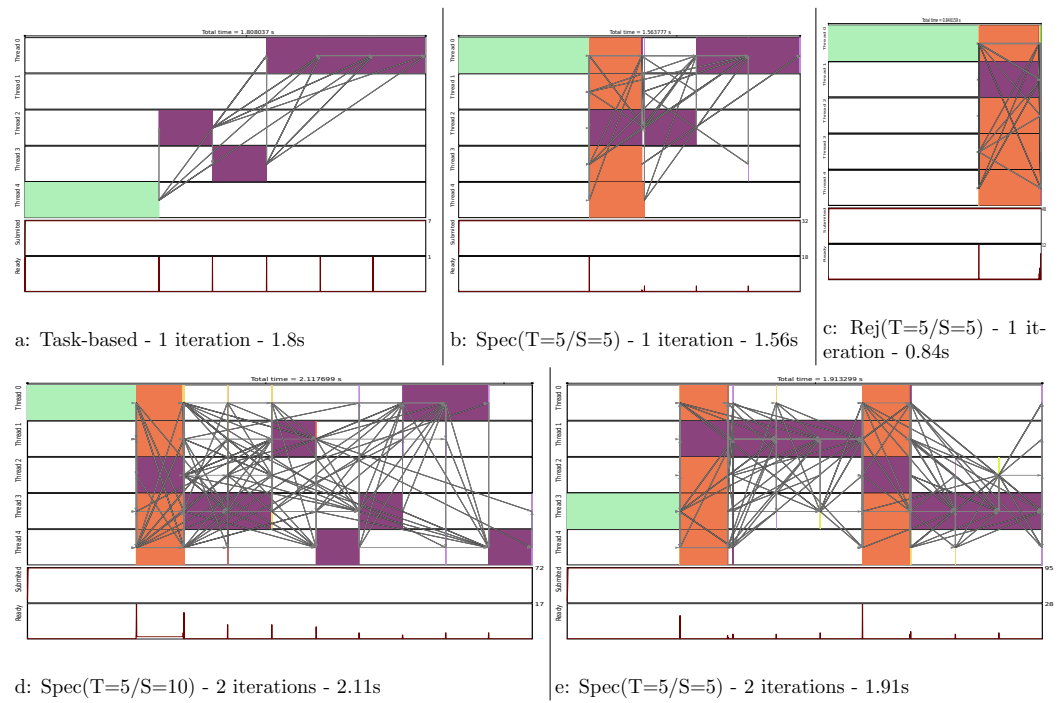

a: Task-based - 1 iteration - 1.8s

b: Spec(T=5/S=5) - 1 iteration - 1.56s

c: Rej(T=5/S=5) - 1 iteration - 0.84s

d: Spec(T=5/S=10) - 2 iterations - 2.11s

e: Spec(T=5/S=5) - 2 iterations - 1.91s

**Figure 11** **Execution traces of the MC simulation for different configurations.** We compute one iteration in (A), (B) and (C). We compute two iterations in (D) and (E). Legend: ■ initial energy computation, ■ move and energy recomputation, and ■ speculative move and energy recomputation.

Full-size 🖼 DOI: 10.7717/peerjcs.183/fig-11

at the performance when all changes are rejected to provide a reference as possible speedup if all the speculations were successful. This is not a realistic execution for an MC simulation because it would mean that all moves are rejected. The speculation is always enabled.

The tasks are created such that they represent one iteration of the loop line 8 in Algorithm 1. They include the move, the computation of the energy and the test for acceptance for a single domain. Consequently, each task accesses in *maybe write* the energy matrix and one of the domains, and in *read* all the other domains.

We show in Fig. 11 the execution trace to compute one or two iterations, where each of the five domains are moved once per iteration. Figure 11A is simply giving the baseline with a task-based execution, but since there is no parallelism we obtain a sequential execution on multiple cores. In Fig. 11B, we see that a normal move/computation task is executed together with four speculative tasks. We can see that the first move was rejected but not the second one. Therefore, the process continued by computing the normal last three moves. In Fig. 11C, we see what would be the execution if all the moves were rejected. In such case, all speculation would be correct and only one normal task would be computed.

In Fig. 11D, we compute two iterations but the $2 \times 5$ uncertain move/computations were inserted consecutively. Consequently, as the second move is rejected, the following eight moves/computations have to be computed normally. To avoid this huge penalty, we can reduce the number of consecutive tasks we insert as shown in Fig. 11E. Here, we restart a new speculative process at each iteration, which limits the degree of parallelization to five (the number of domains) but avoids canceling a complete set of speculative tasks. This means that in the code we manually insert an normal task instead of what could have been a uncertain task to ensure restarting a speculation process afterward.

In Fig. 12A, we provide the performance result for the MC simulation. We look at the gain in performance by using speculation but also at the difference for the speculation degree. We see that as the number of iterations increases, the speedup stabilizes around 40%, which is above the theoretical result for a probability of 0.5 provided in Table 1. Similarly, the upper bound *Rej(T=5/S=5)* gets closer to five.

## REMC

We compare three approaches:

- Task-Based: Since there is no degree of parallelism inside the MC, the maximum parallelism in obtained through the concurrency between replicas.
- Speculative *Spec(T,S)*: Here, *T* represents the number of threads and *S* the number of consecutive uncertain tasks inserted before inserting a normal task. The speculation is always enabled.
- Reject *Rej(T)*: Here, *T* represents the number of threads and *S* the number of consecutive uncertain tasks inserted before inserting a normal task. In this configuration, we look at the performance when all changes are rejected to provide a reference as possible speedup if all the speculations were successful. This is not a realistic execution for an MC simulation. The speculation is always enabled.

Since using speculation creates more work and more tasks to compute, we see that enabling the speculation in all cases could lead to an overhead. This is illustrated in Fig. 13A by the configuration *Spec(T=5/S=4)*. However, using more threads allows better performance, as shown by configurations *Spec(T=10/\*)* and *Spec(T=20/\*)*. In terms of speedup, increasing the number of consecutive speculative tasks does not improve the performance. For instance, *S=2* is always faster because the speculation succeed with a probability of only 0.3 (see Fig. 13B). Of course, we see that having more threads increases the all reject configuration *Rej* performance, and more precisely the more threads we have the faster it executes.

## CONCLUSION

In the current paper, we provided the first results of using speculation in task-based RS's. We described a general pattern and the algorithm of a predictive-oriented approach. Our RS, SPETABARU, is able to execute speculative task flows, and we demonstrate that the obtained speedup for both the MC and REMC is close to the theoretical one of 30%, when the probability to accept changes is 1/2. The mechanism we proposed can be easily incorporated in any other runtime system. However, the fact that we used the *C++* language is an important asset because it makes it possible to generate functions capable of copying

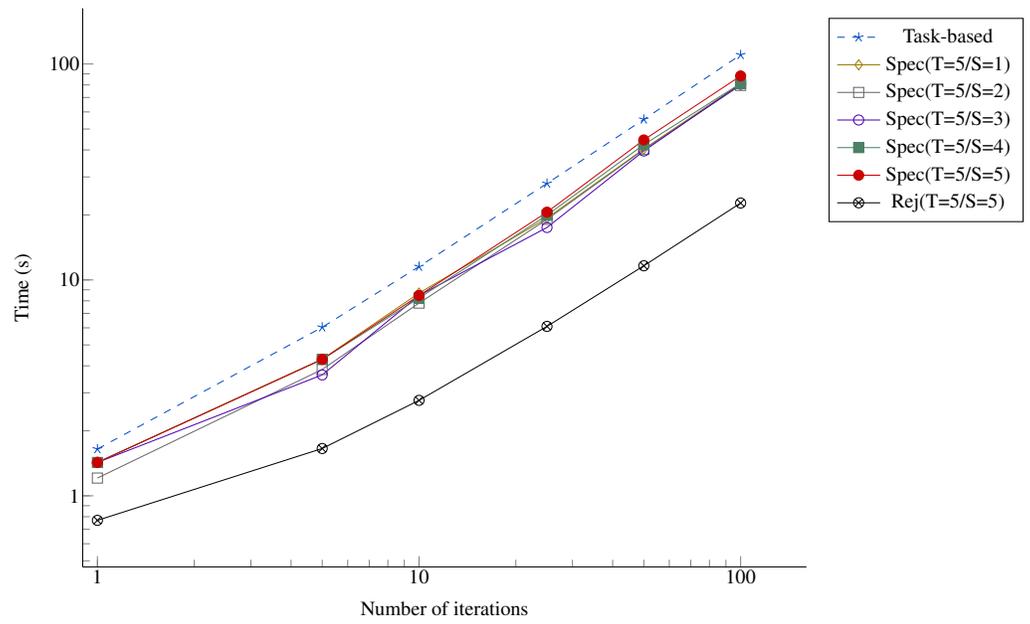

**(a)** Execution time

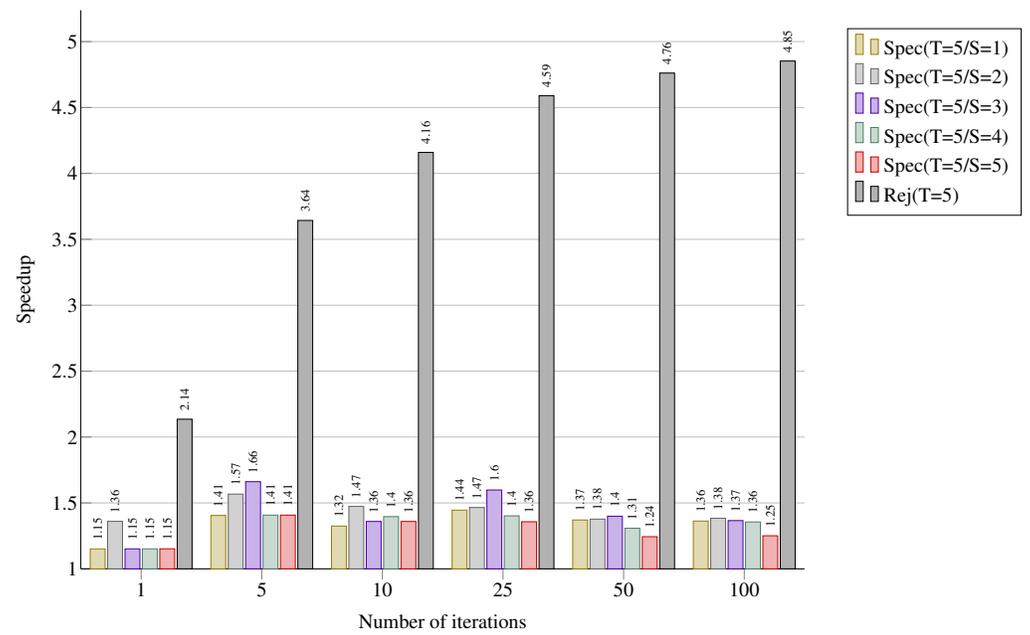

**(b)** Speedup over the *Task-based* executions

**Figure 12** **Performance for the MC simulation with five domains.** (A) Execution time; (B) Speedup over the *Task-based* executions.

Full-size  DOI: 10.7717/peerjcs.183/fig-12

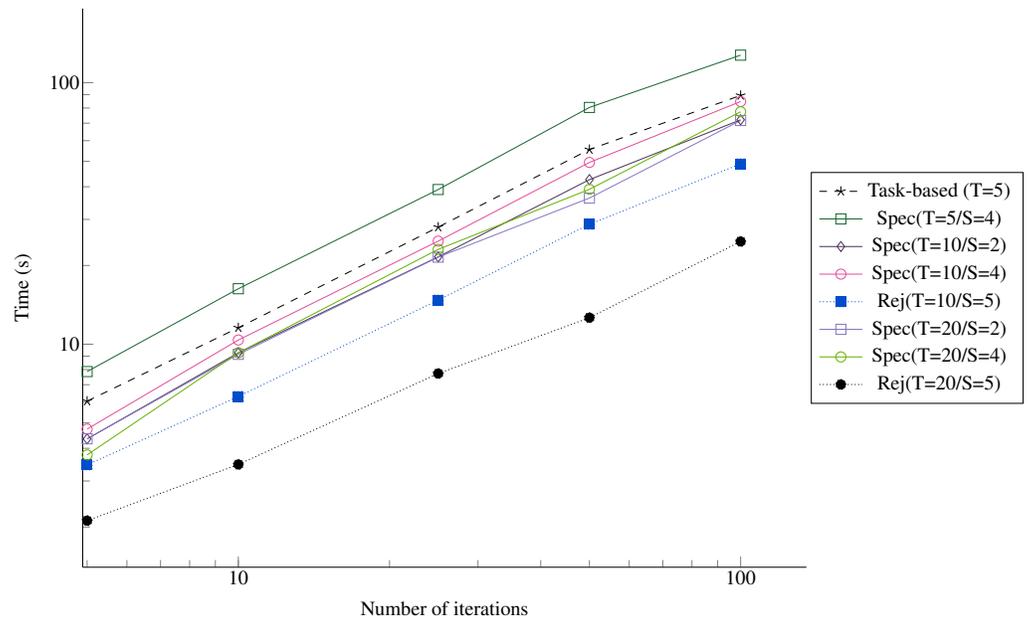

**(a)** Execution time

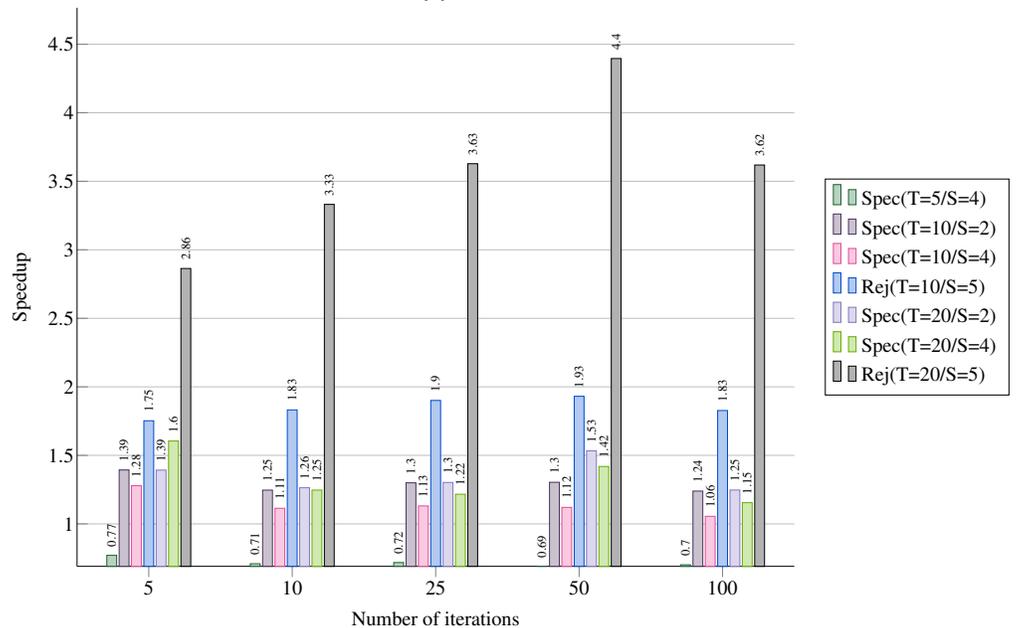

**(b)** Speedup over the *Task-based* executions

(selecting) user's data. The implementation of our system using a low-level language would require using advanced programming from the user perspective such as function pointers.

The number of failures can reduce the benefits and, in the worst case, lead to underperformance when the runtime generates too many tasks for the available CPU cores. In addition, our system uses duplication of data such that memory limits could be reached if the system is not used carefully.

For these reasons, as perspective, we would like to address three major points. First, we would like to automatically limit the degree of speculation (the number of consecutive uncertain tasks). Then, we would like to implement the eager approach to speculate again after a failure in a STG and to obtain a speedup of 2 for MC/REMC simulations. Finally, we would like to study the speculation/decision formula. The resulting process should look at the available ready tasks as the number of threads and certainly use a historical model of the previous execution to predict cleverly if enabling the speculation is appropriate. This will allow removing the possible overhead of the speculation when there is no need to increase the degree of parallelism at some points of the execution.

## ACKNOWLEDGEMENTS

Work by Berenger Bramas was partially done at the Max Planck Computing and Data Facility (MPCDF), Garching, Germany. Experiments presented in this paper were carried out using the PlaFRIM experimental testbed, supported by Inria, CNRS (LABRI and IMB), Université de Bordeaux, Bordeaux INP and Conseil Régional d'Aquitaine

## ADDITIONAL INFORMATION AND DECLARATIONS

### Funding
The authors received no funding for this work.

### Competing Interests
The authors declare there are no competing interests.

### Author Contributions
- Bérenger Bramas conceived and designed the experiments, performed the experiments, analyzed the data, contributed reagents/materials/analysis tools, prepared figures and/or tables, performed the computation work, authored or reviewed drafts of the paper, approved the final draft.

### Data Availability
The following information was supplied regarding data availability:

Data is available at GitHub: https://gitlab.inria.fr/bramas/spetabaru.

### Supplemental Information
Supplemental information for this article can be found online at http://dx.doi.org/10.7717/peerj-cs.183#supplemental-information.